\documentstyle[epsfig]{aipproc}

\begin{document}
\title{Spectral Line Imaging Observations\\of 1E0102.2-7219}

\author{D. S. Davis, K. A. Flanagan, J. C. Houck, C. R. Canizares, 
G. E. Allen, N. S. Schulz, D. Dewey and M. L. Schattenburg}
\address{Massachusetts Institute of Technology\\ Center for Space Research\\
77 Massachusetts Ave\\ Cambridge, MA 02139}

\maketitle

\begin{abstract}
E0102-72 is the second brightest X-ray
source in the Small Magellanic Cloud and the brightest supernova
remnant in the SMC.  We observed this SNR for $\sim$140 ksec with the
High Energy Transmission Gratings (HETG) aboard the $Chandra$ X-ray
Observatory. The small angular size and high surface brightness make
this an excellent target for HETG and we resolve the remnant into
individual lines.  We observe fluxes from several lines which include
O VIII Ly$\alpha$, Ly$\beta$, and O VII along with several lines
from Ne X, Ne IX and Mg XII.  These line ratios provide powerful
constraints on the electron temperature and the ionization age of the
remnant. 
\end{abstract}

1E0102.2-7219 (hereinafter E0102-72) is the brightest X-ray SNR in the
Small Magellanic Cloud (SMC) and was discovered with the Einstein
Observatory's IPC \cite{SM81}. Higher resolution
observations showed that this SNR exhibits a shell-like structure
\cite{Hu88}. Optical emission from E0102-72 was detected
revealing that this is an oxygen-rich SNR \cite{DT81} and
thus the result of the explosion of a massive progenitor. 

A moderate resolution CCD spectrum of this remnant was obtained with
ASCA which observed this object for $\sim$35 ksec. The X-ray spectrum
shows clear lines of oxygen, neon, and magnesium \cite{HK94}.  
Hayashi et al. use the NEI modeling code of Hughes \& Helfand \cite{HH85} 
and find that the data require at least two NEI plasmas.  The 
assumption that each observed element has an independent $nt$ and
temperature lead them to the conclusion that the elements are
not well mixed. 
Hayashi et al. find the NEI parameters for oxygen are log($nt$) =
10.31 s cm$^{-3}$ and log(T) = 7.03~$^\circ$K (0.93 keV) and those for
Ne are log($nt$)=11.37~s~cm$^{-3}$ and log(T) = 6.78~$^\circ$K (0.52
keV).  The fact that the neon has a lower temperature
and higher $nt$ relative to oxygen leads them to the
conclusion that the oxygen emission is related to the forward shock
and the neon to the reverse shock. Our analysis below assumes that 
the emission is from a single $\tau$ plasma model. 

\section*{Chandra HETG Spectra of E0102-72}

The Chandra X-ray Observatory (CXO) observed E0102-72 for a total of
$\sim$140~ksec with the High Energy Transmission Grating Spectrometer
(HETGS) in the optical path of the telescope. The HETGS contains two
types of gratings, the High and Medium Energy Gratings (HEG \& MEG).
These gratings provide an E/$\Delta$E of 100-1000 for point sources
over the energy range of 0.4 - 10 keV.  The HEG and MEG are rotated
with respect to each other so that the dispersed spectra form an ``X''
pattern on the detector. The problem of overlapping orders is resolved
using the moderate energy resolution of the ACIS detector.  Additional 
details of the analysis and results can be found in these proceedings
\cite{can00}\cite{FL00}. Here we present results from the MEG
spectrum of E0102-72. 

We extracted the line fluxes using an annular aperture that encloses
the observed ring. The background was taken from an annular region
outside the dispersed image of interest but with the same center as
the region for the line flux.

\section*{X-ray Morphology}

Figure 1 shows the dispersed image of the SNR in the O VIII
Ly$\alpha$ line (18.97~\AA). While the overall shape of the remnant
can be described as a ring it is clear that this is not true on small
scales. The ring has gaps on the eastern side and a bright knot can be
seen in the southwest quadrant. The image of the SNR in the Ne X
Ly$\alpha$ line (12.13 \AA) is shown in Figure 2. The Ne X image
shows fewer gaps in the rim than the O VIII image, but like the oxygen image
shows deviations from a ring structure. We find no evidence that the
different elements are stratified. However, we do find ionization
structure in the remnant \cite{FL00}.
Despite these complexities a global analysis is useful as it
allows us to determine integrated line fluxes for comparison with 
$ASCA$ and $XMM$ results, and to compare our models with earlier 
results. 


\begin{figure}[ht] 
\begin{minipage}[t]{0.45\textwidth}
\centerline{\epsfig{file=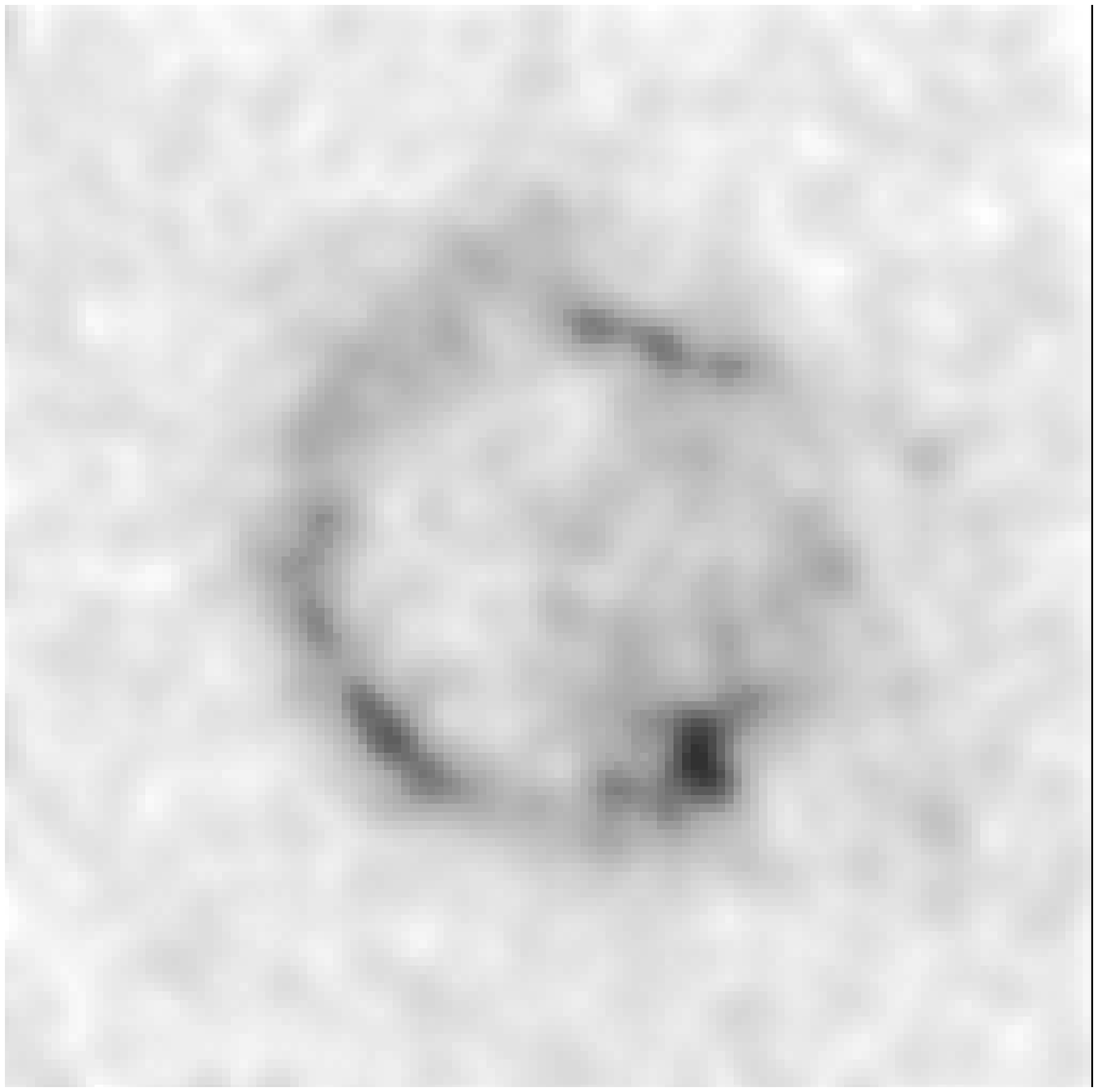,height=2.5in,width=2.5in}}
\vspace{10pt}
\caption{The OVIII Ly$\alpha$ image of E0102-72 from the MEG -1
order. Note the bright knot in the southwest. }
\label{fig1}
\end{minipage}
\hfill
\begin{minipage}[t]{0.45\textwidth}
\centerline{\epsfig{file=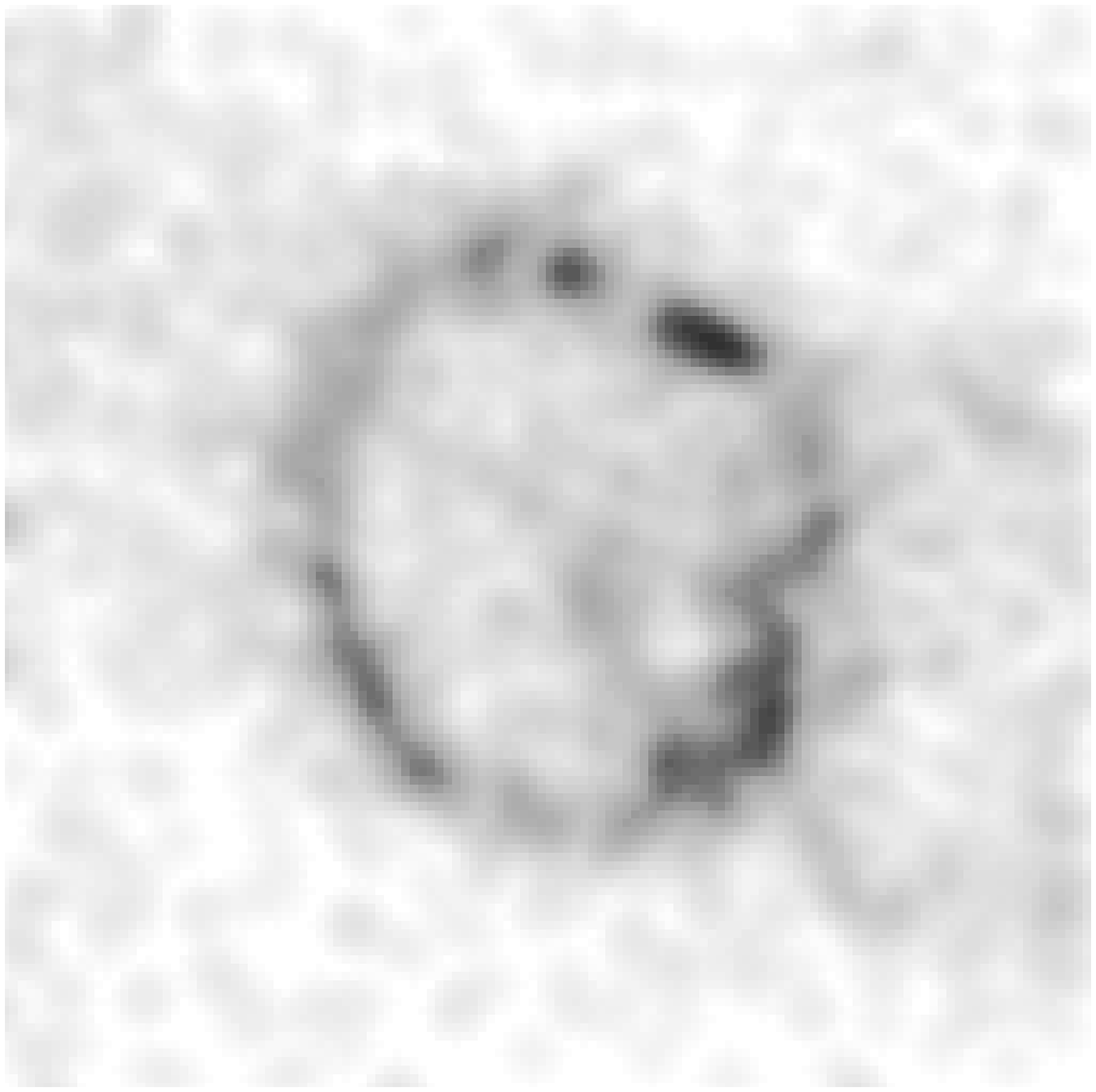,height=2.5in,width=2.5in}}
\vspace{10pt}
\caption{The Ne X Ly$\alpha$ image of E0102-72. This is the data from the 
MEG -1 order.}
\label{fig2}
\end{minipage}
\end{figure}


\section*{Line Diagnostics}

Line ratios were calculated for pairs of hydrogen-like and helium-like
lines and these are compared to the NEI model in XSPEC 11. Here we
present results (Table 1) for the flux ratio of ions from the same element so
that the results are independent of the abundances. We use the column
density of 8.0$\times$10$^{20}$ atoms~cm$^{-2}$ \cite{brdm89} to
correct the flux for interstellar absorption.

We derive tight constrains of the physical state of the plasma from
measurements of O VIII Ly$\beta$/O VIII Ly$\alpha$ and from the ratio
of O VII Forbidden + IC to O VII Resonance. Using Figure 3 we find that the
temperature must be above $\sim$0.68 keV and log($\tau$) is between
9.85 and 10.05 s cm$^{-3}$. The ratios of Ne X Ly$\alpha$ to Ne
IX Res along with Ne X Ly$\beta$/Ne X Ly$\alpha$ also provide tight
constraints on the physical state of 
the plasma. They do not overlap the allowed region for oxygen.  The
allowed region of parameter space is shown in Figure 4; the width of
the regions is determined by the three $\sigma$ limits on the line
ratios. The lower limit on the temperature of the plasma is 0.54 keV
from the oxygen ratio and 0.97 keV from the neon ratio and is
consistent with the $ASCA$ analysis of this remnant \cite{HK94}. The
lower limits for the temperature are robust even for a multi-$\tau$
plasma.

\begin{center}
\begin{tabular}{|l|c|c|} \hline
Line Ratio & Flux Ratio & 3$\sigma$ range\\ \hline
O VIII Ly $\beta$/O VIII Ly$\alpha$ & 0.138 &  0.122 -- 0.154\\ \hline
O VII (For + IC)/ OVII Res           & 0.593 &  0.422 -- 0.764\\ \hline
O VII (4 -$>$ 1)/O VIII Ly$\alpha$  & 0.338 &  0.089 -- 0.587\\ \hline
Ne X Ly$\beta$/Ne X Ly$\alpha$     & 0.141 &  0.124 -- 0.158\\ \hline
Ne X Ly$\alpha$/Ne IX Res          & 0.723 & 0.603 -- 0.843\\ \hline
\end{tabular}
\vskip 0.1cm
{Table 1: The flux ratios for oxygen and neon.}

\end{center}

\begin{figure}
\begin{minipage}[h!]{0.45\linewidth}
\epsfig{file=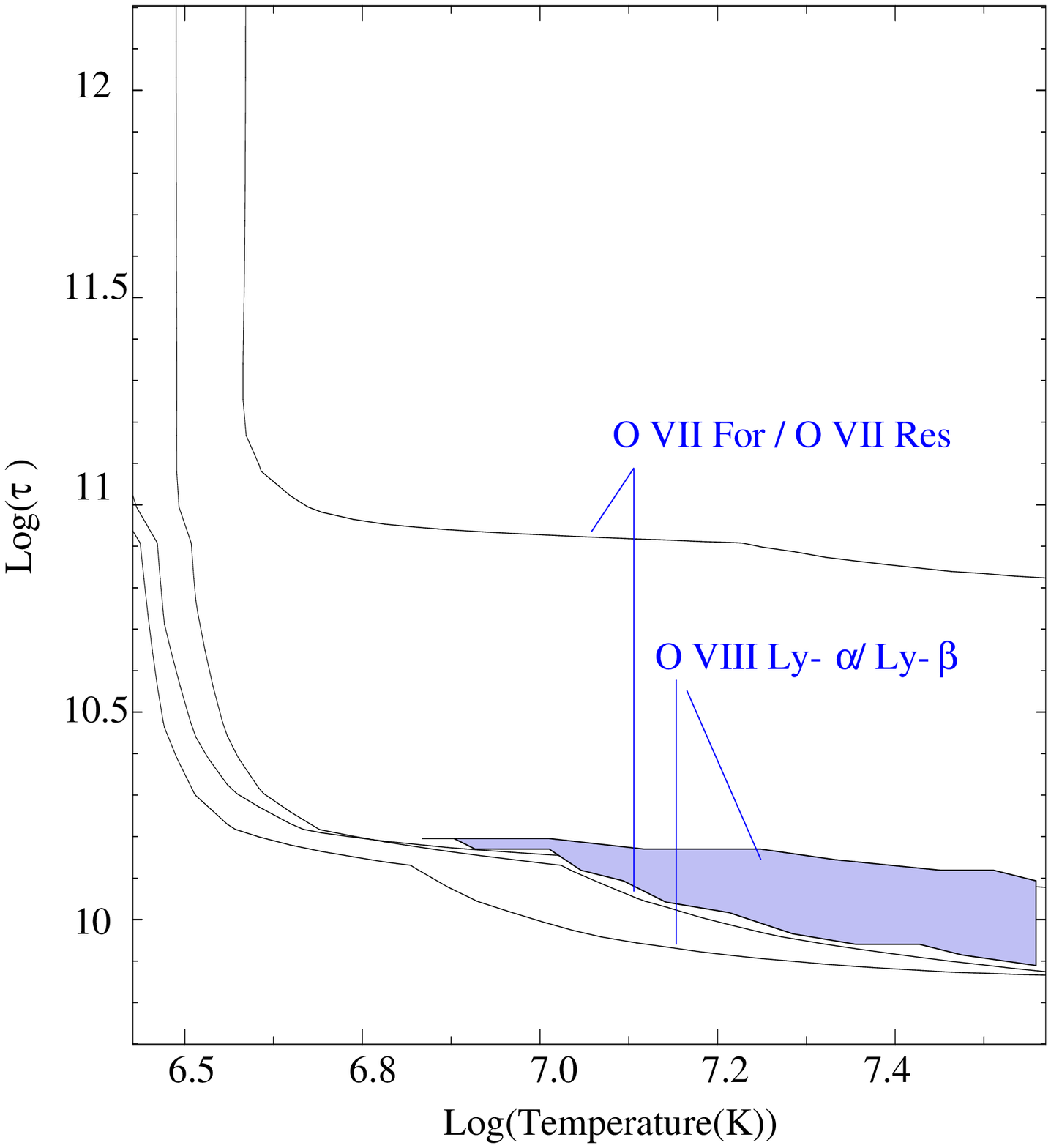,height=2.5in,width=2.1in}
\vspace{10pt}
\caption{The allowed parameter space in $\tau$
Temperature~(K) from the oxygen lines is shown as the shaded region.}
\label{fig3}
\end{minipage}
\hfill
\begin{minipage}[h!]{0.48\linewidth}
\epsfig{file=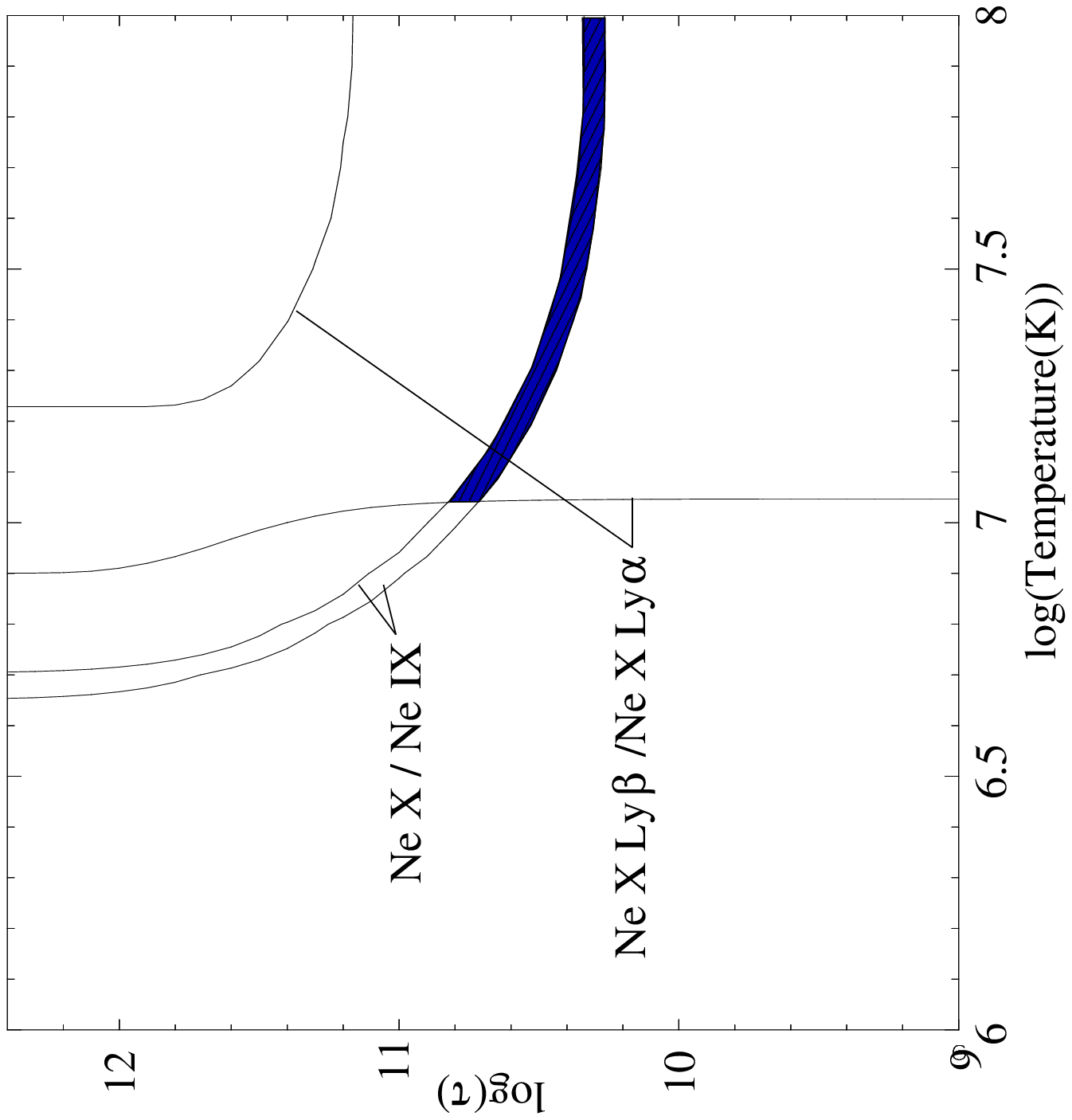,height=2.5in,width=2.1in,angle=-90}
\vspace{10pt}
\caption{The allowed parameter space in $\tau$
Temperature~(K) from the neon lines. Note that the ranges for the x
and y axis are not the same as those in Fig 3.}
\end{minipage}
\end{figure}

\section*{Conclusions}

We have measured the strong emission lines of O and Ne in the SNR
E0102-72. The detected flux ratios from these lines are consistent
with emission from a young SNR that has not yet reached ionization
equilibrium. Our global analysis yields an electron temperature that 
is above $\sim$ 0.54 keV and an ionization age between 
9.85 $<$ log($\tau$) $<$ 10.6, consistent with this being a 
young remnant. 

We find that the line emission from each element,
measured globally, can be described by an NEI plasma using a simple
plane parallel shock model. Our analysis is consistent with the oxygen
and neon emission both being from the reverse shock and we find no
compelling evidence that oxygen and neon are not well mixed. Future
work will include analysis of other lines present in this
spectrum. We also will use the imaging capability of the HETGS to
explore how the plasma diagnostics vary around E0102-72 and map the
temperature and ionization age of this remnant. 

We would like to thank the HETG/CXC group at MIT for their assistance
and useful discussions which contributed to this work. 
This research is funded under NASA contract NAS8-38249 and SAO
SV1-61010

\end{document}